\documentclass{rspublic}
\def\go{\mathrel{\raise.3ex\hbox{$>$}\mkern-14mu\lower0.6ex\hbox{$\sim$}}}
\def\lo{\mathrel{\raise.3ex\hbox{$<$}\mkern-14mu\lower0.6ex\hbox{$\sim$}}}
\begin{document}
\title[Quasar Jets and their Fields]{Quasar Jets and their Fields}
\author[R. D. Blandford]{Roger Blandford}
\affiliation{Caltech, Pasadena, CA 91125, USA}
\label{firstpage}
\maketitle
\begin{abstract}{quasars, accretion discs, magnetic fields, jets}
Observations of jets from quasars and other types of accreting black hole
are briefly summarized. The importance of beaming and
$\gamma$-ray observations
for understanding the origin of these jets is emphasised. It is argued that 
both the power source and the collimation are 
likely to be magnetic in origin, although the details remain controversial.
Ultrarelativistic jets may be formed by the spinning hole and collimated
by a hydromagnetic disc wind.
Progress in understanding jets has been 
handicapped by our inadequate knowledge of how magnetic 
field really behaves under cosmic conditions. Fortunately,
significant insights are coming from solar observations,
numerical simulation and laboratory plasma experiments. 
Some possible, evolutionary ramifications are briefly discussed
and it is suggested that it is the mass of the black hole relative
to that of the galaxy which determines the eventual galaxy morphology.
\end{abstract}
\section{Quasars, Seyferts and Extragalactic
Radio Sources}
\subsection{Quasars}
When the universe was about a quarter 
of its present age, the nucleus
of roughly one out of every five hundred bright galaxies
(at any time) was so bright that it outshone
the surrounding stars. For historical reasons
these nuclei are called quasi-stellar objects 
or quasars for short. 
These quasars are generally recognised by their 
unusually blue optical continua and resulting
broad emission lines. They also are powerful
X-ray sources.  Roughly ten percent of these 
quasars are designated ``radio-loud'', like the first
example discovered, 3C 273,
because they also possess powerful radio sources;
the remainder are ``radio-quiet'', (though
not silent). The separation between these 
two classes is pretty clean, with relatively
few intermediate cases. 
Roughly ten percent of the radio-quiet
quasars also exhibit broad absorption lines
and are called BALQs. (For 
more detailed discussion of much of what follows as well as an extensive
bibliography of original references that cannot be reproduced
here, see \textit{eg} Krolik 1999, Robson 1996.)

Radio-loud quasars are further sub-divided
into ``compact'' and ``extended''
radio sources. The former group are dominated by flat 
spectrum radio nuclei that dominate the emission
at cm wavelengths. The extended sources invariably 
comprise two ``lobes'' of steep spectrum radio
emission straddling the
galaxy and located beyond the observed stars.
We now know that, when observed with greater sensitivity
and particularly at lower frequency, the compact
sources also have extended components,
and, correspondingly, the extended radio sources have
cores that are more prominent at high frequency.
\subsection{Active Galactic Nuclei}
Quasars are really just the brightest members
(with powers in excess of $\sim10^{37}$~W),
of a larger class of ``active galactic nuclei''
or AGN (circumventing some ambiguities).
In fact, it appears that the majority of ``normal''
galaxies exhibit some form of nuclear activity.
One particularly important type of AGN is the 
Seyfert galaxy, first identified 
as a class in the 1940s. Seyfert galaxies come in 
two basic types. Type 1 Seyferts (\textit{eg} NGC 4151)
exhibit both broad and 
narrow optical emission lines and 
poweful soft X-ray emission, whereas Type 2 Seyferts
(\textit{eg} NGC 1068)
only show the narrow lines directly and are
weak soft X-ray sources.
Seyfert galaxies are often considered to be the
low power extension of the quasar luminosity
function, but there are several important 
differences: they are never powerful radio 
sources, they appear to be associated mostly with
spiral galaxies while quasars may reside in giant ellipticals,
Seyfert 1 galaxies are more powerful in X-rays relative
to UV emission and Seyferts never show very broad absorption
lines.
\subsection{Radio galaxies}
Powerful radio sources were first identified with 
giant elliptical galaxies in the 1950s. Like the
quasars they can be extended, (\textit{eg} Cygnus A), or core-dominated,
(\textit{eg} BL Lac, the eponymous ``blazar''). These 
radio sources are supplied with energy,
momentum and magnetic field through a pair of
jets that emerge from a source smaller than the 
compact radio components. The 
extended radio galaxies are further
divided into the weaker ``Fanaroff-Riley'' (or FR) Class 1
objects like Centaurus A and the more powerful
FR2 objects like Cygnus A. It is found that
the brightest radio emitting region is located at
the extremities of the source in FR2 radio galaxies but near
the center of the source in FR1s.
\subsection{Unification}
If all of this sounds a bit confusing, it should.
In fact, the classification of AGN is much more complicated 
than I have described. (The subject is closer in spirit
to clinical psychology than elementary
particle physics!) However,
there has been some progress in bringing order to the field
through a process called ``Unification''.

There are at least four types of unification that 
have been examined. The best established is that 
compact radio sources are extended radio sources
viewed along their relativistic jets.  Essentially
what we are seeing is the relativistically 
Doppler-boosted emission
from the innermost parts of the jet outshining
the unboosted emission from the surrounding extended 
radio source.  (We know that relativistic motions are present
in the compact cores because radio astronomers can image features 
moving across the sky with apparent ``superluminal''
speed.) This explains why, when we observe compact radio
sources at low radio frequency, we see faint, low
surface brightness halos surrounding the compact source.

Almost as well-established is the notion
that Seyfert 2 galaxies are similar to Seyfert 1 
galaxies except that they are observed through
a warped equatorial disc, or torus, that prevents
direct view of the broad emission lines and UV- soft
X-ray spectrum. Here the confirmation is provided by 
detection of broad emission lines from Seyfert 2 galaxies
in polarised 
radiation. This has, presumably, been scattered in 
our direction so as to avoid the disc.

Thirdly, it appears to be the case that many
of the powerful FR2 radio galaxies are actually 
radio-loud quasars that would be classified as
such if we were not viewing them through an obscuring,
dusty gas.

Finally, there is fairly good evidence that most radio-quiet
quasars produce radiatively-driven, equatorial outflows and
we only classify them as BALQ when our line of sight intersects these
flows.
\section{Observations of Black Holes}
\subsection{Black holes as prime movers of nuclear activity.}
Ever since quasars were first discovered in 1963, black holes
have provided one of the most popular explanations for
their activity. (\textit{eg} Lynden-Bell 1969).
They naturally produced 
high radiative efficiency, rapid variation, long-term 
source axes and relativistic 
outflow speeds as the observations required. 
However, it is only in recent years that the positive, observational 
evidence for the presence of black holes in the nuclei
of the majority of regular galaxies
has become overwhelming.  As with stellar-sized black holes,
the only sure approach is dynamical. Both stars and gas have 
had their speeds measured and the combination of speed and size 
suffices to estimate the central 
mass. It has been possible to measure about 25 hole masses.
These range all the way from 
$\sim10^6$~M$_\odot$ to $\sim3\times10^9$~M$_\odot$ and 
have been localised in volumes that, in several cases, are
too small to allow a long-lived star cluster. (The most celebrated 
is now NGC 4258, Moran, these proceedings, which has a mass of 
$3.9\times10^7$~M$_\odot$.) Beyond
all reasonable doubt, these are black holes. In other words, 
this part of the story is on much firmer footing than the rest
of what I shall say.
\subsection{The Galactic center}
A good example is our Galactic centre. Recently,
Ghez \textit{et al} (1998) and Genzel \& Eckart (1997) 
have been able to measure the motion of stars 
close to the hole's location, as determined
by radio astronomy. As they are also able to measure 
the speed of the stars along the line of sight, it will soon
be possible to reconstruct their individual orbits and 
verify directly that the central mass is point-like.  The existing
data are consistent with a black hole of mass $2.6\times10^6$~M$_\odot$
and essentially nothing else.
One important feature of the Galactic centre black hole
is that it is surprisingly under-luminous relative
to the gas supply. Specifically, it seems that
$\sim10^{19}$~kg s$^{-1}$ of gas are supplied
to the black hole. However, the bolometric luminosity
appears to be not much more than $\sim10^{29}$~W
and the quotient gives us $\sim10^{10}$~J kg$^{-1}
\sim10^{-7}c^2$, hardly a good advertisement for gravity
power!  
\subsection{M87}
Another good example is M87, a galaxy in the Virgo cluster.
Here, the speed of the gas orbiting the black hole
is measured and implies a large black hole mass,
$\sim3\times10^9$~M$_\odot$ (see Richstone \textit{et al} 1998
for a general review). Again, both the power
and radiative efficiency are found to be low. M87 
is also a FR1 radio source with a single jet inclined 
at an angle $\sim30^\circ$ to the line of sight.
(There is, presumably, a counter-jet that is rendered invisible
by Doppler beaming.) Despite its low power, features 
have been reported to be moving with apparent speeds $\sim6c$,
perhaps associated with some relativistic gas stream 
deflected slightly closer to the line of sight. 
In a triumph of
precision astronomy, Junor \& Biretta (1995)
have traced this jet to $\sim100m\sim5\times10^{11}$~km
from the hole. This is the best direct
evidence that we have that relativistic jets 
are formed close to black holes.
\subsection{Spin}
The mass $m$ of a black hole, expressed in 
geometrical units, just determines a scale
of length and time.  Of more physical 
interest is a second
parameter which measures the shape of the surrounding
spacetime. It is convenient to choose this to be the spin
angular velocity of the hole, $\Omega$.  This
has a maximum value $1/2m$, corresponding to an
extreme Kerr hole. The most convincing case,
presented to date for having measured
this quantity has been given for MCG 6-30-15
(Tanaka \textit{et al} 1995) where the shape of the 
measured $\sim7$~keV Fe line profile is similar
to that expected to be produced by 
an accretion disc extending 
down to its limiting, least stable, circular orbit
from a hole that is spinning nearly maximally.
Unfortunately
this is not the only interpretation of this 
profile (Reynolds \& Begelman 1997). 
\subsection{X-ray binaries}
Although these are not quasars, there are at least
nine X-ray binaries where the mass of the compact
object exceeds the maximum mass of a neutron star
or white dwarf. Particularly prominent among these
objects are the microquasars, discussed here by Mirabel.
Several of these sources exhibit ``quasi-periodic
oscillations'' or QPOs, presumably originating 
from short-lived disc modes. The frequencies of these
modes must measure the hole mass and spin,
though it has not yet been possible to explain how in
a convincing manner.
\section{Observations of Jets}
\subsection{The $\gamma$-ray revolution}
The general existence of jets, similar to those previously
observed in M87 and 3C273, was inferred in extragalactic, double 
radio sources from the demonstrated need for 
a continuous supply of energy and linear momentum (Rees 1971).
Bipolar outflows are now known to be a common
feature of accreting objects, specifically, they have also been found 
in association with microquasars, young stellar objects and
neutron star binaries. Until about eight years ago, this subject was the 
almost exclusive province of the radio astronomer.
However, with the success of the EGRET instrument 
on Compton Gamma Ray Observatory, (Hartmann \textit{et al} 1999),
it has become apparent that the radio emission
is often, and probably always, a bolometrically insignificant 
part of the jet luminosity. (The great strength of radio 
observations is that they enable us to image jets directly in fine detail.)

Extragalactic jets (\textit{eg} Hughes \textit{et al} 1991,
Ostrowski \textit{et al} 1997) present the cleanest
challenge to astrophysicists. Let us draw together the evidence
from several well-studied black holes sources and gamble that 
they are fundamentally similar structures.  We can then 
formulate a general model of jet formation
and collimation. From the M87 observation, it 
appears that jets are formed as
collimated ultrarelativistic outflows on scales smaller than 
$\sim100m$. Their initial composition is not known,
but they quickly become prodigious emitters of GeV $\gamma$-rays
and are variable on time scales as short as $\sim30m$
(Wehrle \textit{et al} 1998). If we 
correct for relativistic kinematics, then the GeV $\gamma$-ray
emission region is probably
located at radii $\lo10^{3-4}m$. 

An important, energy-dependent radius 
is that of the ``$\gamma$-sphere'',
where the optical depth for a $\gamma$-ray to create
pairs by combining with X-rays is unity
(Blandford \& Levinson 1995). 
A second important radius is the ``annihilation radius'',
within which electrons and positrons can cool
to subrelativistic energies and annihilate in one expansion
time. If efficient particle acceleration occurs between the annihilation
radius and the  $\sim0.5$~MeV $\gamma$-sphere, 
then the jet is likely to comprise 
electron-positron pairs at larger radii. (Whether
or not there is evidence for this is an interesting
controversy at the present time, Wardle \textit{et al} 1998.)
However, the inner jet cannot comprise only pairs.
There must be a second agency to carry the momentum
and to overcome radiative drag so that the jet can persist 
to larger radii, as observed. The two candidates are 
electromagnetic field and protons, with the former being
preferred because any scheme to create a directed proton beam 
would probably require invoking an even larger electromagnetic
energy density.

The $\gamma$-ray spectrum extends up to $\go$~1TeV 
(Quinn \textit{et al} 1996) and
is produced by inverse Compton scattering of soft photons
that are probably created within the jet by the synchrotron
process, although they may also be part of the ambient
radiation field. (TeV sources are rapidly 
variable but can only be observed out to 
$z\lo0.1$, because of absorption on the cosmic
infrared background.) The inner jet must therefore be capable of 
accelerating electrons to energies $\go1$~TeV. It is hard to estimate
accurately the jet beaming factor and efficiency and the amount 
of obscuration, but it appears that a typical ultrarelativistic jet 
carries a time-averaged power that is significant
fraction, perhaps a few percent, of the total emitted power of 
the underlying quasar or AGN. In a few cases, (\textit{eg} Cygnus A),
this fraction may be more than a half.
\subsection{Radio observations}
At larger jet radii, $\sim10^{5-6}$m, the outflow is essentially
adiabatic. Initially the radio synchrotron emission is
self-absorbed and unresolved - the radio core.
However, at larger radii, the emission is optically thin and radio
astronomers are able to track the motion of relativistic
shocks, the superluminally expanding features, travelling along 
the jet accelerating high energy electrons as they go.
Optical and X-ray synchrotron emission is observed 
out to quite large radii ($\sim10^7m$) in a few sources 
(\textit{eg} M87), implying that these shocks are capable 
of accelerating $\sim100$~TeV electrons.

We know enough about the physical conditions
in the radio-emitting regions to place a lower bound on the 
internal pressure and to compare this with the maximum external
gas pressure at the same radius, deduced on the basis of 
X-ray observations. In the most powerful sources, the jet appears
to be overpressured by factors $\sim10-100$.
This disparity provides one of the strongest reasons for invoking
magnetic collimation and confinement of jets.

A major uncertainty in our understanding of jets is 
the bulk Lorentz factor of the outflow, $\Gamma$. This is 
important because the observed flux density from a coherent 
feature moving towards us increases $\propto\Gamma^3$, at
the transformed frequency and so a relatively insignificant
part of a poorly collimated outflow can outshine all of the 
rest of the jet for selected observers. (Note that $\Gamma$ 
refers to the motion of the emitting material,
not the motion of the peak of the emission. In a shock, these
are distinct.) Furthermore, although
it has long been thought that the observation of superluminal
expansion speeds $\lo10c$ suggested that $\Gamma\lo10$,
typically, we are now beginning to suspect
that values an order of magnitude higher may be present.
This is because of the discovery of intraday variability
of the cm emission in several sources. If this
is intrinsic to a synchrotron source, it implies that $\Gamma\go1000$
in some cases and requires unreasonably large jet powers
(Kedziera-Chudzczer \textit{etal} 1997). A more 
reasonable explanation of these variations is refractive
scintillation in our interstellar medium. However, there 
may still be a problem because scintillation
cannot change the source brightness temperature which 
has been measured to be as large as $\sim10^{14}$~K
two orders of magnitude in excess of the 
value needed to match the inverse Compton emission to the
synchrotron emission.
As brightness temperatures are boosted by one power
of the Doppler factor in relativistic expansion,
this suggests that $\Gamma\go100$.  A second 
reason for considering larger Lorentz factors is that
the precedent has already been set by models
of the most energetic $\gamma$-ray bursts which suggest
that they are beamed towards us with $\Gamma\sim300$.
All of this discussion has prompted a re-examination
of coherent emission mechanisms which are not subject
to these constraints, though are subject to other
limitations.
\subsection{What do we know about jet magnetic fields?}
Radio polarisation observations indicate that the 
magnetic field in a jet is relatively ordered. 
On arcsecond scales, a characteristic pattern is observed with the 
more powerful FR2 radio galaxies exhibiting a parallel
magnetic field while the less powerful FR1 source show 
perpendicular magnetic field, (although parallel 
fields are sometimes seen at small radii and at the outer
edges of resolved jets). The interpretation is 
straightforward. The magnetic field 
direction reflects the rate of shear
of the velocity field, with the parallel case arising when there
is a significant velocity gradient across the jet or in a boundary
layer and the perpendicular field predominating 
when the transverse expansion is most important, that is to say
after internal velocity gradients have been erased. Note that we are 
supposing that the \textit{mean} field component
along the jet is very small.
This is reasonable on general grounds because the associated 
magnetic flux would otherwise be very large and, as it is only likely
to decrease along the jet, it would be associated with an
unreasonable magnetic pressure close to the black hole.

There are also patterns on the milliarcsecond scales probed 
by VLBI observations (\textit{eg} Gabuzda \textit{et al} 1999).
These show similar patterns to observations 
on larger angular scales and are also consistent with 
emission from travelling internal shocks.
\subsection{Do discs launch hydromagnetic jets and, if so, how?}
Having described the magnetic field that we observe directly,
what about the field that we cannot see?  There are
three distinct processes to which it may be contributing:
launching of a powerful outflow close to the hole, 
collimation of this outflow into the narrow jets observed at somewhat larger
distance and confinement of these jets on all scales out to the extended radio
components. We have already attributed confinement on VLBI scales
to magnetic stress.
Can invisible magnetic field lines, initially frozen into a highly
conducting accretion disc close to the black hole, 
wrapped around the jet, and each other, 
by the differential rotation do the rest of the job?

Several general collimation mechanisms have been described (\textit{cf}
Pudritz, these proceedings, \textit{eg} 
Mestel 1998, K\"onigl \& Pudritz 1999 and references therein.) In particular,
if there is a strong, ordered magnetic field, the tension associated 
with its azimuthal component 
creates a collimating and confining ``hoop'' stress. The gradient 
in the magnetic pressure may help. Magnetic field attached 
to an accretion disc also
provides a means of launching a jet because it
will exert a torque on the disc and extract some of 
the binding energy released by the infalling gas. A hydromagnetic
wind may also remove a significant fraction of the accreting
mass because, if the field direction
subtends an angle of more than $\pm30^\circ$ to the rotation
axis, then gas will be flung away from the disc by centrifugal force.
The resulting, collimated, hydromagnetic outflow is likely to have
an asymptotic speed a few times the escape velocity 
at the magnetic footpoint on the disc. This elementary mechanism
is straightforward to describe and analyse using similarity methods
(\textit{eg} Ostriker 1997).

Another type of ordered MHD outflow model has been developed by
Shu \textit{et al} (1994), more specifically for application 
to young stellar objects (\textit{cf} Pringle, these proceedings).
Here, it is proposed that essentially all of the magnetic 
flux in the magnetosphere emanates from the innermost radii
of the disc. One concern with this model is that some of these
field lines must lie on the surface of the accretion disc
and be subject to rapid reconnection as they sweep by loops 
and prominences attached to the disc. If this happens, the flux will
migrate radially outward very quickly.

An organised field need not be unipolar. Lynden-Bell (1996) has 
developed quasi-static, force-free models in which a poloidal magnetic loop 
is twisted by a differentially rotating disk and rapidly 
expands outwards creating toroidal field of one sign.
One possible problem with this mechanism is that it is assumed
that the Alfv\'en speed $v_A$ is infinite, whereas, in practice,
matter is likely to be flung out as well 
so as to lower the Alfv\'en speed 
below the rotational speed close to the disk
and the outflow velocity far from the disk.
However, the foot points of a given loop will be
differentially rotating and 
although the field line attached to the inner
footpoint trails, the field line at the outer footpoint leads.
Now a leading field line looses causal contact with the disk above a height
$\sim v_A/\Omega$ where it must be dragged by the returning 
field from the inner footpoint. The whole loop must therefore
be sub-Alfv\'enic. Loops that become super-Alfv\'enic
are presumably unstable and will shock, reconnect and detach. 

Magnetic field can also be responsible for powering jets in a less organised
manner. As discussed here by Brandenburg, 
the magnetorotational instability drives a non-helical
dynamo (Balbus \& Hawley 1998) and
ensures that accretion discs are 
able to regenerate radial and toroidal magnetic field on an orbital
timescale and build up an internal disc field that is supposed
to be much stronger than any ordered, vertical field that leaves the disc
surface. Under these circumstances, loops of field of size comparable
with the disc thickness (or more relevantly, the pressure
scale height) will be continuously released by buoyancy and reconnection
from the disc surface into an active corona in much the same way
as is envisaged to happen at the surfaces of stars and the Galactic disc
(Miller \& Stone 1999 and references therein). Not only 
does this process seem unavoidable, it also provides a suitable
power source for the X-ray emission of Seyfert galaxies and other AGN.  
However, it does not automatically lead to a collimated outflow. 
Tout \& Pringle (1996) have suggested that these small loops 
can grow through an inverse cascade to a size
$\sim r$ and that these larger loops can 
provide enough tension to effect collimation despite the 
reversals in the sign of the azimuthal field with 
cylindrical radius and the isolation from the underlying
rotating disk. In an alternative description,
Heinz \& Begelman (1999) have suggested that the field is
disordered on small scales and that its dynamical effect
may be approximated by a local mean stress tensor.
In a hybrid model (Blandford \& Payne 1982, Emmering 
\textit{et al} 1992), the rms coronal field
is supposed to be larger than the mean vertical 
field and constantly changing on timescales shorter than an orbital period
through reconnection. This allows matter to be injected, 
intermittently, onto open field
lines and flung out into a wind where the field becomes 
relatively organised and smoothly varying. 

A key difference in approach underlying these
models is whether the mean field at high altitude
is unipolar and established through a balance between 
advection by the inflowing disk from large radius
(where most of the flux resides) and escape 
through reconnection or whether it is created locally by dynamo
action so that the horizontal correlation length is $\lo r$.
What happens is unclear.
On the one hand, van Ballegooijen (1989) argued that the advection
rate is slower than the escape rate
by a factor $O(r/HPr_m)$, where $H$ is the disc thickness
and $Pr_m$ is the effective, magnetic Prandtl number.
Therefore if $Pr_m<r/H$, and it is traditionally set to unity
in a turbulent medium, then any large scale field must escape.
Alternatively, we can express the mean inflow rate in the disk
as $t_{{\rm in}}^{-1}\sim\alpha\Omega(H/r)^2$, where
$\alpha\sim0.1$ is the assumed coefficient of proportionality 
between the shear stress and the pressure, and observe that,
to within a numerical factor of order unity, 
this is the rate at which magnetic field will random walk 
out of the disk by reconnecting on flux loops of size
$\sim H$ every $\Omega^{-1}$. 

Of course, the polarity of the field 
associated with the accreting gas is likely to change. This does 
not preclude confinement by the hoop stress associated with the 
toroidal field, for which the polarity must also change. To see
this, consider an elementary model in which an axial current $I$ 
flows along a jet.  There is an axisymmetric, 
toroidal field of strength $\mu_0I/2\pi r$. Now let there
be axisymmetric, axial currents of strength $2I$ and alternating sign 
flowing within thin cylindrical sheaths of radius $r_1,r_2,\dots$. 
The toroidal field magnitude will be unchanged but the sign will reverse.
The stress acting on the current sheets $\mu_0I^2/8\pi^2r_i^2$,
will be balanced across them and will steadily 
decrease until it can be matched onto
an ambient gas pressure. Stress balance within the current
sheets must be achieved either with gas pressure or a rotating 
magnetic field. This configuration is presumably
tearing mode unstable and magnetic energy will be steadily dissipated
through reconnection. However, it should persist for long 
enough in a super-Alfv\'enic outflow to allow jet collimation.   

As must be clear by now, this is a contentious subject. As has also 
been true of cosmology, there
are several different elementary models
that are amenable to applied mathematical analysis
without any guarantee that their underlying assumptions are relevant to 
the application. We simply do not understand MHD well enough yet
to know what are the correct assumptions to use and this is a
pre-requisite to answering the big astrophysical questions.
We need to know the ratio of the internal torque
transporting angular momentum radially outward through the disk
to the external torque applied to the disk surface
and responsible for carrying off angular momentum in 
the wind. (Note that if the mean vertical field threading 
the disc is large, then the magnetorotational instability is likely to be 
suppressed.) Furthermore, we want to understand the magnetic structure 
and energy balance of the disk corona. Presumably it is a low $\beta$ plasma
which can be approximated as force-free, in contrast to the disk field. 
Stability is another issue. For example, simple prescriptions for specifying
the rate at which mass is loaded onto open field
lines lead to the conclusion that a centrifugally-driven
wind is unstable (Lubow \textit{et al} 1994); 
alternative prescriptions lead to stationary, self-adjusting 
flows (K\"onigl \& Wardle 1996, Krasnopolsky 1999, in preparation). 
Another difficult issue is the nature of the boundary conditions
to apply at large and small cylindrical radius. In a similarity solution
the difficulty is finessed. However, in  a finite disk the ultimate collimation
can be strongly influenced by what is assumed (Okamoto 1999 and references therein). 

However, the situation is not hopeless.  There are at least three lines
of inquiry that are helping a lot. Before I explain why, though, I would 
like to discuss two further, relevant questions.
\section{Is Adiabatic Accretion Conservative?}
In recent years, there has been renewed interest in 
what happens when gas accretes at a slow rate or, more
specifically at low density (relative to Eddington
accretion). Under these circumstances there is the 
possibility that the flow is adiabatic (in the sense
that it does not cool on a dynamical timescale). This can surely happen if
the only coupling between
the ions and the electrons is through Coulomb scattering. 
When the mass accretion rate
in units of the Eddington rate, ($4\pi GM/c\kappa_T$), denoted $\dot m$ is 
$\go0.3\alpha^2$, then cooling will be
ineffective and the disc is likely to inflate as a consequence of its large
internal energy. There has been a lot of work in recent years describing
conservative flows (called ADAFs, \textit{eg} Narayan,
Mahadevan \& Quataert 1998, Kato \textit{et al}
1998 and references therein) that carry all the 
supplied mass down the black hole
with negligible radiative loss. Fitting the emissivity computed
from these flows to diverse, observed sources has been a relatively
successful enterprise.

However, there are some fundamental, dynamical problems
with ADAF solutions. The gas is formally unbound, mainly because the 
viscous torque that allows them to proceed, transports energy as well as
angular angular momentum and this must be dissipated in a differentially
rotating disc. To be specific, if there is an extensive, adiabatic,
conservative, viscous disc flow, then it can be shown that
the specific energy of the gas is twice its orbital kinetic energy. 
The model, as it stands, does not appear to be self-consistent 
without becoming quasi-spherical and then the gas close to the rotation 
axis is unsupported. 

For these and other reasons, Blandford \& Begelman (1999) have proposed
that adiabatic accretion always be accompanied by outflows which 
carry off energy, angular momentum, mass etc in unspecified
amounts that are sufficient to allow the gas to accrete
on bound orbits. The outflows may be gas dynamical
or hydromagnetic.  In these ``ADIOS'' solutions, the rate at which 
gas actually accretes onto the black hole can be orders of magnitude
less than the rate at which it is supplied. If this view of adiabatic 
accretion ultimately prevails, and there are some ways by which it can be 
distinguished observationally from ADAF flows, then there will
be a good dynamical reason why accretion is often accompanied by outflow.  

A second mode of adiabatic accretion can occur
when the gas is supplied at a rate far in excess of the 
Eddington rate ($\dot m\go10$). Under these circumstances, there is no difficulty
in emitting radiation. The problem arises when the photons try to 
escape (Begelman \& Meier 1982). It turns out 
that they will be trapped in the accreting gas
and advected in toward the hole faster than they can diffuse away.
Again the flow is likely to be effectively adiabatic and is likely
to drive an outflow for the same reason as a sub-critical inflow.
If this view turns out to be correct, it will be hard for black holes
to accrete mass at a rate that is much larger than roughly ten times
the Eddington rate. These outflows, launched initially by Thomson 
scattering and then further accelerated by resonance line scattering
may be associated with the absorbing gas in BALQs. 
\section{Are Quasar Jets Powered by Black Hole Spin Energy?}
When a black hole spins and its
spacetime is described by the Kerr metric, a fraction,
$\le0.29$ of its mass energy can be associated 
with its spin and is extractable. A gedanken experiment
to do just this was performed by Penrose (1969). 
For the $\sim 3\times10^9$~M$_\odot
\equiv5\times10^{56}$~J, black hole in M87, perhaps $\sim10^{56}$~J
of energy can realistically have been tapped over its life
which is ample to account for an extremely profligate youth.
\subsection{How to get Blood out of a Stone}
The most natural way to tap this energy is by using large scale magnetic
field (Blandford \& Znajek 1977, Thorne \textit{et al}
1986, Lee, Wijers \& Brown 1999). Currents flowing in the inner accretion disc can support
a significant amount of magnetic flux (typically $\Phi\sim10^{25}$~Wb
threading the hole.  Now the hole
can be considered to be a good, though not perfect, electrical 
conductor, with a surface conductance of 
$(E/H)_{{\rm horizon}}=Z_0=377\Omega$. Therefore,
when the hole spins, it can act as a unipolar inductor and create an emf 
${\cal E}\sim\Omega\Phi\sim10^{20}$~V, (just about 
sufficient to accelerate the UHE cosmic rays). 
This emf can drive a closed, field-aligned 
current circuit that dissipates both within the 
horizon of the hole (the internal resistance of the circuit) 
and in the particle acceleration region at the base of the
jet (the load). As the total resistance  of the circuit is
$\sim100\Omega$, the current, in this example, is $\sim10^{18}$~A
and the power $\sim10^{38}$~W. The
power can be thought of as being transported 
away from the horizon in the form of a Poynting flux.
(The ``no-hair'' theorem is not violated because the flux of energy 
is only conserved in the frame that is not-rotating with respect
to infinity. Observers that hover just above the horizon
must rotate with respect to this frame 
and they would observe an inwardly-directed
energy flux.) The electromagnetic power scales according
to the memorable relation, $L\propto a^2B^2c$, where $B$ is the 
field that \textit{threads} the hole.  (This stipulation is important because
magnetic flux is unable to penetrate the horizon when the rotation 
rate is nearly maximal and so the electromagnetic power is reduced). 

This mechanism has recently attracted attention because of its possible
role in powering $\gamma$-ray bursts. 
The magnetic field is quite likely to be separated from the 
accreting gas so that the resulting outflow can move ultrarelativistically.
Similar, though less extreme, conditions are required in 
quasar jets. However, it has also been argued that the power 
extracted from the hole is likely to be much less than that
extracted hydromagnetically from the inner disc, mainly because the area
of the latter is larger (Livio, Ogilvie \& Pringle 1999). 
This is probably true, at least for a thin
disk (\textit{cf} Blandford \& Znajek 1977). However,
quasar jet powers are only a fraction 
of the bolometric power, as a comparison of the $\gamma$-ray background
with the quasar light background affirms and they can still have a black hole
origin. Furthermore, in a thick disk, perhaps associated with an 
adiabatic inflow, a funnel can form and the jet power fraction
can plausibly become quite large (\textit{eg} Rees \textit{et al} 1982).

Another objection has been put forward by Natarajan 
\& Pringle (1998) who have presented a new analysis
of the Bardeen-Petterson mechanism whereby a spinning hole will
interact dynamically with a misaligned outer disk.
They conclude that black holes will align more rapidly
than previously estimated and, if the plane of the gas 
supply keeps changing, the hole will spin down faster than
accretion will cause it to spin up. (Note that this alignment
is coupled with a quite large release of energy in the outer disk 
$O(m^3\Omega\omega_{{\rm BP}}\theta^2)$, where 
$\omega_{{\rm BP}}$ is the Keplerian angular velocity at the warp radius
and $\theta$ is the misalignment between the hole spin
and the outer disk angular momentum.) The estimated alignment timescales
typically fall between the jet transit times to the outer lobes of 
radio sources and the overall radio source lifetimes consistent with the
``dentist's drill'' model of lobe advance. VLBI observations
are already resolving scales $\sim r_{{\rm BP}}$ and may soon determine
if the jet axis is determined by the hole or the disc. 

Despite all these concerns, there is still a particularly good reason for invoking
the extraction of electromagnetic energy from a spinning hole to power 
quasar and similar jets.  This is because, as I have emphasized,
they are ultrarelativistic and, initially, probably magnetically-dominated.
The hydromagnetic winds from the surface
of a disc are unlikely to avoid being loaded with plasma and, consequently,
will be unlikely to achieve high terminal Lorentz factors. No such
drawback attends the field lines that thread the surface of a black hole!
\subsection{Flogging a Dead Horse}
In a variant of this mechanism, that has also just 
been resurrected, it may be possible to transfer angular momentum
from the hole directly to the disc. 
Krolik (1999) and Gammie (1999) have considered magnetic field in the 
plane of the disc within the innermost stable circular orbit and 
shown that it is possible for magnetic torque to carry an energy flux 
outward along a magnetic flux tube, at least outside the
ingoing Alfv\'en critical point and, in this manner, increase
the specific energy release from the disc to a 
value above that associated with the binding energy at the 
innermost stable circular orbit. The extra power
is, again, derived from the spin of the hole.  One 
particular concern about this mechanism is that the 
infalling gas remain magnetically attached to the disc. It seems quite 
likely that both senses of radial field will be present within the 
transition region between the disc and the hole and that 
magnetic reconnection will happen quite freely. An alternative
way to model this interaction (Blandford \& Spruit,
in preparation), is to suppose that magnetic flux tubes
connect the disc surface to the event horizon at intermediate 
latitude, extracting energy and angular momentum from the hole. 
Note that, in this case, the magnetic field is tied to 
the orbiting gas not the central 
object - the opposite of what happens with accreting neutron stars.
\section{Laboratories for Studying MHD}
As I have already emphasized, we are seriously handicapped by our
ignorance of how magnetic field actually behaves in cosmic environments.
Fortunately there are three promising approaches to 
improving our understanding.
\subsection{Cosmic laboratories}
The first of these is direct observation of dynamical magnetic fields.
\subsubsection{Solar wind}
Observations of the solar photosphere and corona by YOHKOH, SOHO and TRACE,
as reported here by Title, have led to a revolution in our general
perspective on MHD. No longer is the observed as having a bland surface 
occasionally ruptured by coronal arches. Instead, it is a tightly interwoven,
largely invisible tapestry of moving magnetic field, 
energised by the underlying
convection and held together for as long (typically 1-2~d)
as reconnection and intercommutation of magnetic field lines 
can be staved off. The observed 
corona is maintained at a temperature of $\sim1-2\times10^6$~K,
probably by reconnecting nanoflares, though the detailed
calorimetry is still a matter of controversy. The 
magnetic field dominates the energy density. By contrast,
$\beta>1$ below the photosphere, where it has just been found that 
a dynamo may be operating, in addition to the dynamo located at the
base of the convection zone that produces the field of sunspots.

Observations during solar minimum 
of the high latitude wind at $\sim1-2$~AU by Ulysses have been no less
instructive (\textit{eg} Fisk 1998).  
The wind and its associated flux emanate from 
giant coronal holes located near the poles. The radial magnetic field at 1~AU has 
a steady value of about 4~nT and satisfies an inverse square variation with radius
indicating that it more or less fills most of a hemisphere in a uniform manner.
The total field pattern executes a Parker spiral, although it is 
slightly over-wound at the poles.
The density is $\sim3\times10^{-21}$~kg m$^{-3}$ and again is pretty  constant
with time and latitude, satisfying an inverse square variation with radius
suggesting that the surprisingly large  and uniform 
outflow speed of $\sim800$~km s$^{-1}$
is also achieved well before $\sim1$~AU.

The wind cannot be thermally-driven; instead,
it is supposedly energised by Alfv\'en waves close to the corona. 
We can actually use this information to estimate the torque that the solar wind
exerts upon the sun at this time. If we approximate the wind as spherical and the 
radial velocity as pretty constant, then the torque is given by
\begin{equation}
G\sim2\pi r^4v_r\rho\Omega\sim\frac{2\pi B_r^2r^4\Omega}{\mu_0v_r}
\end{equation}
evaluated at the Alfv\'en radius, where $v_r=B_r(\mu_0\rho)^{-1/2}$.
This is given by $G\sim10^{23}$~Nm and the Alfv\'en radius evaluates
to $\sim12$~R$_\odot$. The solar angular momentum is, for comparison,
$\sim1.6\times10^{41}$~kg m$^2$~s$^{-1}$ and so the characteristic slowing 
down time is now $\sim30$~Gyr, consistent with expectation. 

A more sophisticated comparison is certainly possible and would
have to take into account that the coronal hole axis is inclined
with respect to the spin axis, that there is a separate slow wind at low 
latitude, that there is a significant amount of Alfv\'en turbulence,
that the sun differentially rotates, that there is a solar cycle and so on
(\textit{cf} Fisk 1996). However
even though the sun is slow rotator relative to
an accretion disk, it clearly has much to teach us as we try
to model hydromagnetic winds from accretion disks. In particular
it may already have told us that simple, stationary, axisymmetric 
models of disk winds are a good starting point.
\subsubsection{Crab Nebula}
Recent HST observations of the Crab Nebula also have some lessons for us.
The main reason is that a similar path is being followed by the 
energy: ordered rotational energy $\rightarrow$ electromagnetic energy
$\rightarrow$ relativistic electron energy $\rightarrow$ non-thermal radiation.
Moving features have been observed associated with the
famous wisps, which may coincide with the strong relativistic
shock formed where the momentum flux in the outflow matches the 
ambient nebula pressure (Gallant \& Arons 1994). Our understanding
remains somewhat sketchy, but this is the closest we are ever likely 
to be to particle acceleration in an ultrarelativistic plasma,
so it is worth persisting.
It ought to be possible to understand the speed and composition
of the outflow and whether or not a strong collisionless shock
is really formed. The absence of a narrow jet may well point to the 
importance of having an extended disc for forming jets.
\subsubsection{X-ray astronomy}
The next eight months, will see the launch of three complementary X-ray telescopes 
Chandra, XMM and ASTRO-E. They should improve our understanding
of the structure of jets, discs and the cosmological distribution of quasars 
in much the same way that YOHKOH, SOHO and TRACE have so enriched our view of the sun.
\subsection{Numerical MHD laboratories}
The second laboratory is computational. As we have seen from Brandenburg's
talk (and his cited references), the capability to 
perform relatively high dynamic 
range, three dimensional numerical MHD is already here 
and there have already been serious attempts to expand this 
capability into the realm of special and general
relativity. This facilitates a variety of numerical experiments.  For example
Stone \textit{et al} (1999) have recently carried out a series
of two dimensional 
simulations of adiabatic accretion in which they consider 
purely hydrodynamical flows and introduce a variety of \textit{ad hoc}
prescriptions for the viscosity. They find that the gas becomes 
strongly convective with the mean specific angular
momentum being constant on isentropes and the mean mass flow through 
radius $r$ settling down to a non-conservative variation,
$\dot M\propto r$, consistent with the predictions of a limiting ADIOS solution.
However, instead of creating a supersonic wind, the outflow is subsonic
and is mostly confined to the surface of the disc. This highlights that
an extra source of entropy must be present at the disc
surface for a disc to create a supersonic outflow purely hydrodynamically.
Furthermore, by contrasting these simulations with 
their hydromagnetic counterparts (Hawley, 1999, in preparation), 
it has become clear that the 
dissipation associated with the magnetic torque must be handled 
very carefully numerically and that the character of the flow may depend upon what
is assumed.
For example, the magnetic field may create a turbulence 
spectrum that is absorbed roughly volumetrically
at some inner scale through transit 
time damping (Gruzinov 1998, Quataert 1998), or
the entropy may be produced in a small fraction of the total volume
if reconnection at current sheets dominates (Bisnovatyi-Kogan \& Lovelace 1997). 
Although in both cases the dissipation is local,
the latter assumption is likely to lead to higher temperatures and a 
different emissivity than the former. Alternatively,
most of the energy may be transported hydromagnetically from the disc
to the corona so that there is little local dissipation.
These assumptions may lead to quite different flows.
\subsection{Plasma physics laboratories}
We have already mentioned several instances where 
magnetic reconnection can have a major role in determining 
the energy release and the details of the flow.  As 
described here by Parnell, we still do not have a confident understanding
of this important process and some novel reconnection modes are 
currently under serious consideration. The way reconnection works in two
dimensions is now
well understood and the emphasis has shifted towards studying ways in 
which it may operate in three dimensions (\textit{eg} Priest \& Titov 1996,
Galsgaard \& Nordlund 1997). Another way to 
approach this problem is through laboratory experimentation.
Although the magnetic Reynolds' numbers are never as high in 
the laboratory as one would like, it is still possible to perform 
instructive experiments and then to scale with the 
relevant dimensionless numbers (\textit{eg} Brown \textit{et al} 1998)
so as to learn how astrophysical reconnection occurs under differing
conditions.
\section{What Next?}
In this review, I have tried to consider the problem of 
understanding powerful, relativistic, quasar jets in a general context.
An outline of one solution, in which an electromagnetic core
is collimated by a non-relativistic, hydromagnetic wind, has existed
for twenty years. However, it has not been satisfactorily
verified observationally and there are several, genuine 
physical difficulties that are not understood. There is still 
a chance that a quite different and essentially
non-magnetic mechanism might be at work. However, as I 
have emphasised, the powerful 
combination of numerical simulation and direct observation
of ``real'' plasma is forcing us to become much 
more sophisticated and further
important insights are likely over the next few years.

I would like to conclude by mentioning some speculative
extensions of this model to a ``grand unified theory'' of AGN
that attempts to give a comprehensive interpretation of all 
of the principal modes of nuclear activity that are observed.
This is stimulated by two recent observational claims.
Firstly, Magorrian \textit{et al} (1998)
have argued that the black hole mass in local, dormant galaxies
is proportional to the mass of the ``bulge''.
(Ellipticals are all bulge; spirals have progressively
smaller bulges as the type changes from Sa to Sd.)
Secondly, McLure \textit{et al} (1999) have presented evidence that 
quasars are surrounded by elliptical galaxies, not spirals
as once thought.

Suppose that black holes grow, as argued above, at an
Eddington-limited rate early in the life of a galaxy.
The rate of mass supply should decline with time, whereas
the Eddington limit grows with mass. It is possible that the hole
will grow with an e-folding timescale
$\sim30$~Myr until it reaches a mass somewhat below its
present value. When the mass supply is super-Eddington, 
and the hole mass exceeds $\sim10^8$~M$_\odot$, the object
will form a radio-quiet quasar and produce a high speed,
radiatively-driven wind.
During the final e-folding of hole mass, this wind
will deposit $\sim10^{53}(m/10^8{\rm M}_\odot)$~J of energy in the outer parts
of the galaxy and, presumably, will drive a blast wave into the infalling gas. 
If we assume that the escape velocity is $\sim300$~km s$^{-1}$ up to 
$10^{12}(m/10^8{\rm M}_\odot)$~M$_\odot$ of gas can be driven away.
Allowing for inefficiency and radiative loss, 
there is enough energy to expel the gas
and forestall the formation of a disk if $m\go10^8$~M$_\odot$. 
In other words, relatively big black holes lead to elliptical galaxies.
We know that BAL outflows are not seen when the 
luminosity is less than that associated with a quasar. (The 
explanation may be a subtle effect associated with opacity.)
It is then possible that smaller holes cannot expel the infalling gas
and that a disc will form. In other words galaxies form around 
black holes, not \textit{vice versa}. 

One key observation that must still be explained is 
that low mass holes / spirals / Seyfert 
galaxies do \textit{not} create powerful, ultrarelativistic jets.  
Perhaps accretion continues for much longer at
an intermediate rate as the supply of gas is not shut down
and this is sufficient
to drive the spin of the hole to nearly its maximal value, (as reported
for MCG 6-30-15). This will prevent the hole from 
forming a powerful, relativistic jet. Alternatively,
the collimating hydromagnetic disc wind may just not be produced
at an intermediate accretion rate.
By contrast, with a high mass hole / elliptical / quasar, the mass supply
may quickly decline below Eddington and perhaps become adiabatic
close to the hole. A radio-loud quasar or FR2 radio galaxy 
can then form. This will persist while the spin energy is 
depleted and the central jet becomes progressively 
less powerful. Eventually, the jet thrust becomes less than 
that associated with the disc wind and, although the observed 
jet may be formed with speed $\sim c$, it will soon be decelerated
by interacting with the more slowly moving wind. This is an FR1 radio galaxy.
If nothing else happens, the jet and nuclear activity will
finally decline to dormancy. However, if two old galaxies and their black holes
subsequently merge, a fairly rapidly spinning hole may ensue 
and a powerful radio source will be re-born (Wilson \& Colbert 1995). 

These speculations have clear, observable implications.
\begin{acknowledgements}
I thank the Royal Society for support and the other attendees for instruction and 
suggestions. I also thank participants in the Black Hole Astrophysics 
Program at the Institute for Theoretical
Physics, Santa Barbara, notably, for many discussions 
of these and related topics. This research was also supported 
by the Beverly and Raymond Sackler Foundation,
NSF grants AST 95-29170, 99-00866, PHY 94-07194 and NASA grant 5-2837.
\end{acknowledgements}

\label{lastpage}
\end{document}